\documentclass[12pt, notitlepage]{article}
\pdfoutput=1

\usepackage{float}
\usepackage{subfig}
\usepackage{graphicx}
\usepackage[T1]{fontenc}
\usepackage[utf8]{inputenc}
\usepackage{authblk}
\usepackage{amsmath,bm}
\usepackage[left=3cm,right=3cm,top=2cm,bottom=2cm]{geometry}
\renewcommand\footnotemark{}
\begin{document}

\title{Hypothesis on the Nature of Time}         
\author{D.N.~Coumbe}
\affil{\small{\emph{Faculty of Physics, Astronomy and Applied Computer Science, Jagiellonian University, ul. prof. Stanislawa Lojasiewicza 11, Krakow, PL 30-348}}\footnote{E-mail: daniel.coumbe@uj.edu.pl}}        
\date{\small({Dated: \today})}          
\maketitle


\begin{abstract}

We present numerical evidence that fictitious diffusing particles in the causal dynamical triangulation (CDT) approach to quantum gravity exceed the speed of light on small distance scales. We argue this superluminal behaviour is responsible for the appearance of dimensional reduction in the spectral dimension. By axiomatically enforcing a scale invariant speed of light we show that time must dilate as a function of relative scale, just as it does as a function of relative velocity. By calculating the Hausdorff dimension of CDT diffusion paths we present a seemingly equivalent dual description in terms of a scale dependent Wick rotation of the metric. Such a modification to the nature of time may also have relevance for other approaches to quantum gravity. 

\vspace{1cm}

\end{abstract}


\begin{section}{Introduction}

General relativity and quantum mechanics, in their respective domains of applicability, are extremely accurate theories. However, when strong gravitational fields interact over short distances, such as in the vicinity of the big bang singularity or near black holes, the description of such phenomena demand a unification of the two theories; a theory of quantum gravity.
  
The lack of experimental data at Planckian scales has allowed a surplus of observationally indistinguishable approaches to quantum gravity to coexist, yielding numerous often conflicting predictions about the nature of space, time and matter. Given the diverse number of different approaches to quantum gravity, perhaps it is prudent to look for common features shared by all such theories. One seemingly ubiquitous feature of quantum gravitational theories is the thermodynamic behaviour of black holes, which appears so consistently across such a diverse number of approaches it seems likely that it will feature in some form or other in the correct version of quantum gravity. 

Another feature common to nearly all approaches to quantum gravity is dimensional reduction. A number of independent field-theoretic approaches to quantum gravity, using a variety of different techniques, have reported that the dimension of spacetime appears to be scale-dependent. Causal dynamical triangulations (CDT) \cite{Ambjorn:2005db,Coumbe:2014noa}, exact renormalisation group methods \cite{Lauscher:2005qz}, Ho{\v r}ava-Lifshitz gravity \cite{Horava:2009if}, noncommutative geometry \cite{Arzano:2014jfa,Benedetti:2008gu}, loop quantum gravity \cite{Modesto:2008jz} and string theory \cite{Atick:1988si,Calcagni:2013eua} have all reported that the dimension of spacetime appears to reduce as one probes spacetime on ever decreasing distance scales, or conversely with ever increasing energies. Individually these results do not constitute substantial evidence in support of dimensional reduction; collectively, however, they form a compelling argument that demands further attention.\interfootnotelinepenalty=10000 \footnote{\scriptsize We refer the reader to Refs. \cite{Carlip:2009kf,Carlip:2011tt} for an excellent review of the accumulating evidence for the appearance of dimensional reduction, as well as an alternative proposal to explain the appearance of dimensional reduction on Planckian scales.} 

\begin{subsection}{Motivations for challenging the reality of dimensional reduction}\label{motivations}

If the dimension of spacetime really does decrease as we probe spacetime on small distance scales then we must accept the seemingly unphysical consequences; that superluminal motion is possible \cite{Sotiriou:2011aa,Amelino-Camelia:2013tla}, that relativistic symmetries are at the very least deformed \cite{Amelino-Camelia:2013cfa,AmelinoCamelia:2000mn,KowalskiGlikman:2001gp,Magueijo:2002am}, that it is possible to break Lorentz invariance \cite{Sotiriou:2011mu}, that gravitational waves cannot exist in lower dimensional formulations of general relativity \cite{Gott:1986bp} and that Maxwell's theory of electromagnetism completely breaks down \cite{Weyl1922}. In light of these radical consequences, coupled with the absence of any solid theoretical explanation underpinning the mechanism of dimensional reduction, we are motivated to challenge the physical reality of this phenomenon and question what the appearance of dimensional reduction is really telling us. Although the central focus of this work is CDT quantum gravity, due to the almost universal observation of dimensional reduction in different approaches to quantum gravity it is possible that the arguments used and the conclusions reached in this work may be more widely applicable.

The spectral dimension $D_{S}$ is a measure of the effective dimension of a manifold over varying length scales (see e.g. \cite{Ambjorn05} for a thorough discussion), and is related to the probability $P_{r}$ that a random walk will return to the origin after $\sigma$ diffusion steps. The spectral dimension is defined by

\begin{equation}
D_{S}=-2\frac{\rm{d}\rm{log}P_{r}}{\rm{d}\rm{log}\sigma}.
\end{equation}

A spectral dimension that varies with distance scale implies either a systematic violation, a non-systematic violation or a deformation of Lorentz invariance. Systematic violations of Lorentz invariance are position independent and result in the existence of a preferred global reference frame \cite{Mattingly:2005re}. Non-systematic violations of Lorentz invariance are when the symmetry varies stochastically as a function of position \cite{Mattingly:2005re}. Deformations of Lorentz invariance occur when the exact low-energy symmetry is deformed, but not violated, in the high-energy limit \cite{Kostelecky02}. Regardless of whether Lorentzian symmetry is violated or just deformed by a varying spectral dimension, one will nevertheless obtain a dispersion relation $E(p)$ that is modified to some extent \cite{Kostelecky02}.

The dispersion relation

\begin{equation}
E^{2}=p^2\left(1+\left(\lambda p\right)^{2\gamma}\right),
\label{sol0}
\end{equation}

\noindent is an example of a systematic violation of Lorentz invariance, as it implies a global choice of frame in which energy is a distinguished component of momentum \cite{Mattingly:2005re,Amelino-Camelia:2013tla}. For the majority of approaches to quantum gravity a scale dependent spectral dimension may give only a non-systematically violating or deformed dispersion relation, as most approaches do not allow for the possibility of a preferred global reference frame. However, there exist a small number of special cases that allow for the possibility of a global preferred frame of reference, such as Ho{\v r}ava-Lifshitz gravity and possibly CDT\interfootnotelinepenalty=10000 \footnote{\scriptsize The question of whether the fixing of the foliation in CDT is at odds with general covariance remains an open issue. If CDT is consistent with general covariance, it should be possible to obtain the same results using an EDT formulation, which is explicitly covariant from the outset. However, this does not currently appear to be the case in 4-dimensions \cite{Coumbe:2014nea}. Also the similarity between CDT and Ho{\v r}ava-Lifshitz gravity, which is known to have a preferred global frame, is suggestive \cite{Ambjorn:2010hu,Ambjorn:2013joa}. There exist a number of lower-dimensional counter-examples that contend that CDT does not require a preferred frame, see e.g. \cite{Jordan:2013awa,Markopoulou:2004jz}.}, for which it is then possible that a reduction of the spectral dimension may imply a dispersion relation of the type of Eq. (\ref{sol0}), but of course it may equally well lead to a non-systematically violating or deformed dispersion relation. Keeping in mind that this particular dispersion relation is only a possible example, and is not expected to be typical, we nevertheless note that Eq. (\ref{sol0}) implies a modified speed of light that is given by

\begin{equation}
c_{m}=\frac{E}{p}=\sqrt{1+\left(\lambda p\right)^{2\gamma}}.
\label{sol1}
\end{equation}

Nearly all approaches that report dimensional reduction find a large distance value for the spectral dimension of 4, and a small distance value of 2. In approaches to quantum gravity for which Eq. (\ref{sol1}) might be applicable, the small distance dimensionality of 2 can be translated into a $\gamma$ value of 2 \cite{Amelino-Camelia:2013tla}, thus giving a speed of light $c_{m}$ that is dependent on the momentum scale $\lambda p$. In fact, as is evident from Eq. (\ref{sol1}), any non-zero value of $\gamma$ results in a speed of light that is dependent on the momentum scale $\lambda p$. In a spacetime with $(d_{H}+t_{H})$ Hausdorff dimensions one finds the general form of the spectral dimension \cite{Amelino-Camelia:2013tla,Horava:2009if} 

\begin{equation}
D_{S}=t_{H}+\frac{d_{H}}{1+\gamma}.
\label{hausSpeed}
\end{equation}

In Ref. \cite{Sotiriou:2011aa} an expression for the ratio of the phase and group velocity $v_{phase}/v_{group}$ in terms of the spectral dimension is derived in $(d+1)$ topological dimensions, independent of any particular approach to quantum gravity.\interfootnotelinepenalty=10000 \footnote{\scriptsize Equation (\ref{GroupVel}) is based on a saddlepoint approximation (see Ref. \cite{Sotiriou:2011aa} for details).}

\begin{equation}
D_{S}=1+d\frac{v_{phase}}{v_{group}} + ...
\label{GroupVel}
\end{equation}

\noindent For electromagnetic waves in a vacuum one should obtain $v_{group}/v_{phase}=1$, which we take to define a dimensionless speed of light parameter $c_{m}$. However, for any degree of dimensional reduction $D_{S}<4$, Eq. (\ref{GroupVel}) clearly indicates that the dimensionless value for the speed of light $c_{m}=v_{group}/v_{phase}$ must exceed unity.\interfootnotelinepenalty=10000 \footnote{\scriptsize Obviously the phase velocity of light can in certain circumstances exceed $c$, for example when travelling through certain dispersive media, but the signal velocity is strictly forbidden from doing so. In a vacuum, electromagnetic waves must obey the relation $v_{group}/v_{phase}=1$, a relation that is violated by dimensional reduction.} 

From these two examples we see that dimensional reduction suggests the speed of light can be scale dependent. Additionally, any theory with a running dimensionality must at the very least deform the relativistic symmetries \cite{Amelino-Camelia:2013cfa,AmelinoCamelia:2000mn,KowalskiGlikman:2001gp,Magueijo:2002am}, and may not preserve Lorentz invariance \cite{Sotiriou:2011mu}, although in certain cases it may still preserve the relativity principle \cite{Amelino-Camelia:2013cfa}. The more radical stance would be to interpret dimensional reduction as support for variable speed of light theories (VSL) \cite{Moffat:1992ud,Albrecht:1998ir}, or theories that explicitly break Lorentz invariance such as Ho{\v r}ava-Lifshitz gravity \cite{Horava:2009uw}. However, the much more conservative stance is to explicitly preserve the constancy of the speed of light from the outset, and compute the consequences. 

\end{subsection}
\end{section}


\begin{section}{A scale dependent speed of light in CDT}\label{cdtderiv}

The canonical point in the physical phase of CDT, which has an established macroscopic 4-dimensional de Sitter geometry \cite{Ambjorn:2008wc}, has been shown to have a scale dependent spectral dimension given by the functional form 

\begin{equation}
D_{S}=a-\frac{b}{c+\sigma}.
\label{funcform}
\end{equation}

\noindent Where $a$, $b$ and $c$ are free fit parameters, and $\sigma$ is the diffusion time. CDT simulations yield a fit to the data with $a=4.02$, $b=119$ and $c=54$ \cite{Ambjorn:2005db}. A more recent study at the same point in the CDT parameter space also gives similar results, namely $a=4.06$, $b=135$ and $c=67$ \cite{Coumbe:2014noa}. This particular functional form of the spectral dimension was also arrived at using purely analytic considerations in Ref. \cite{Giasemidis:2012qk,Giasemidis:2012rf}. Integration of Eq. (\ref{funcform}) gives an expression for the probability $P_{r}$ that the diffusion process will return to the origin after $\sigma$ steps

\begin{equation}
P_{r}=\frac{1}{\sigma^{a/2}\left(1+\frac{c}{\sigma}\right)^{\frac{b}{2c}}}.
\end{equation}

\noindent As found in Refs. \cite{Ambjorn:2005db,Coumbe:2014noa} $a\approx 4$ and the ratio $b/2c\approx 1$, hence to a good approximation one finds\footnote{\scriptsize This expression is also equivalent to the continuum return probability guessed at in Ref. \cite{Ambjorn:2005db} upon equating $c=Al_{p}^{2}$, where $l_{p}$ is the Planck length and $A$ is a numerical constant.} 

\begin{equation}
P\left(\sigma\right)\approx\frac{1}{\sigma^{2}+c \sigma}.\interfootnotelinepenalty=10000 
\label{RetProb2}
\end{equation}

The probability of return for infinitely flat 4-dimensional Euclidean space with no dimensional reduction is given by $P\left(\sigma\right)=\sigma^{-2}$. The path length traced out by a diffusing particle is just proportional to the number of diffusion steps $\sigma$. We ask what function $\Gamma$ rescales this path length $\sigma$ such that we obtain the probability of return found in CDT, namely that of Eq. (\ref{RetProb2}). Hence we form the equation 

\begin{equation}
\frac{1}{\Gamma^{2}\left(\sigma\right)\sigma^{2}}=\frac{1}{\sigma^{2}+c \sigma},
\end{equation}

\noindent which gives a $\Gamma$ function of  

\begin{equation}
\Gamma\left(\sigma\right)=\sqrt{1+\frac{c}{\sigma}}.
\label{GammaFuncSig}
\end{equation}

\noindent Thus, by rescaling $\sigma$ in the expression for infinitely flat 4-dimensional Euclidean space by the $\Gamma$ function of Eq. (\ref{GammaFuncSig}) we obtain the probability of return found in CDT, and hence upon taking the logarithmic derivative we obtain the functional form of dimensional reduction reported in CDT \cite{Ambjorn:2005db,Coumbe:2014noa}. Since $\sigma$ is proportional to the square of the distance scale $\Delta x$ with which one probes the manifold, we can write this $\Gamma$ function as

\begin{equation}
\Gamma\left(\Delta x\right)=\sqrt{1+\frac{c}{\Delta x^{2}}}.
\label{GammaFuncX}
\end{equation}

\noindent Assuming the free fit parameter $c$ in Eq. (\ref{GammaFuncX}) can be expressed in Planck units by $c=Al_{p}^{2}$ as first suggested in Ref. \cite{Ambjorn:2005db}, where $l_{p}$ is the Planck length and $A$ is a numerical constant, then Eq. (\ref{GammaFuncX}) becomes

\begin{equation}
\Gamma(\Delta x)=\sqrt{1+\frac{Al_{p}^{2}}{\Delta x^2}}.
\end{equation}

By substituting the fit to the functional form $D_{S}=a-b/(c+\sigma)$ found using the CDT approach to quantum gravity into Eq. (\ref{GroupVel}) we obtain a modified speed of light $c_{m}$, as implied by dimensional reduction in CDT, 

\begin{equation} 
c_{m}=\frac{v_{group}}{v_{phase}}=\frac{d}{a-\frac{b}{c+\sigma}-1}.                                                                                            
\label{cm}
\end{equation}
                                                                                                                                                                                
\noindent Figure (\ref{Gamma3}) shows this modified speed of light $c_{m}$ as a function of $\sigma$, with the fit parameters $a=4.06$, $b=135$ and $c=67$ determined from CDT calculations \cite{Coumbe:2014noa}.\interfootnotelinepenalty=10000 \footnote{\scriptsize Ref. \cite{Mielczarek:2015cja} provides an independent derivation of superluminality in CDT that is almost identical to the one presented in this work.} 


\begin{figure}[H]
\centering
\includegraphics[width=0.7\linewidth,natwidth=610,natheight=642]{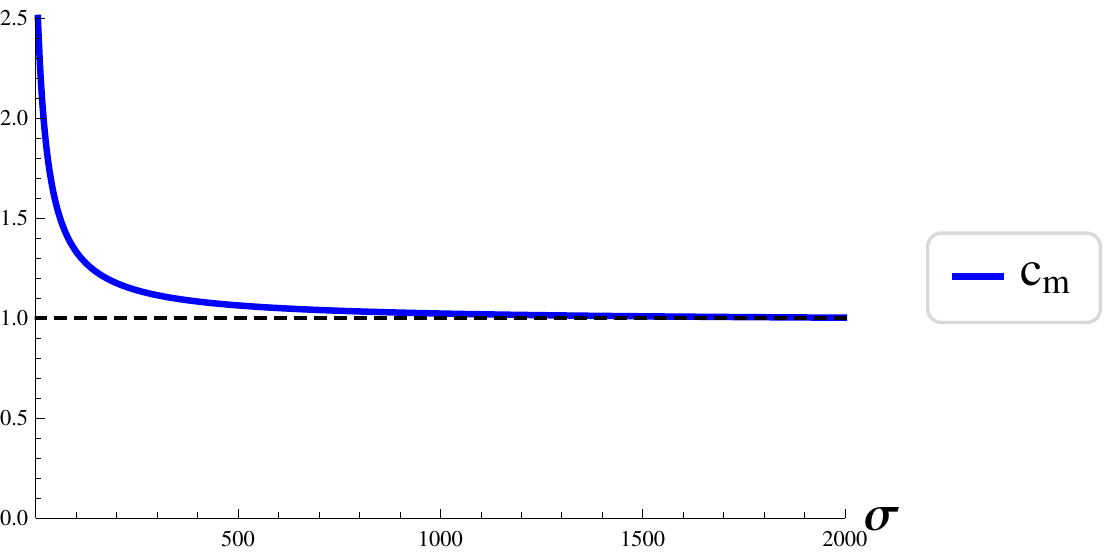}
\caption{\small The scale dependent speed of light $c_{m}$ as predicted by dimensional reduction in CDT.}
\label{Gamma3}
\end{figure}



As can be seen in Fig. (\ref{Gamma3}) the modified speed of light $c_{m}$ increases above unity as one probes the manifold on ever decreasing distance scales. This kind of variable speed of light seems to be a generic feature of theories containing dimensional reduction. We attribute this to the distance the diffusing particle travels dilating as we probe the manifold on smaller scales, as is common to all fractal curves. To get an intuitive understanding for why we expect to see an apparently decreasing dimension, and hence a variable speed of light, if length dilates on smaller distance scales consider the following example. Imagine a diffusion process on a one-dimensional line, at each point on the line the diffusing particle has just two options; it can either go left or right (assuming the particle must move). For a two-dimensional diffusion process on a grid, at each point the particle has four options. In three-dimensions the particle will have six options, etc. Since having more options decreases the probability of returning to the origin, it is clear that dimension is inversely related to the probability of return. Hence, in order to explain why the dimension appears to reduce on small scales, and by extension why the speed of light increases, one only has to explain a relative increase in the probability of return on smaller scales compared with larger ones. Now, if the path length of the diffusing particle increases on smaller distances the probability that the diffusing particle will return to the origin will be comparatively increased, because for a given number of diffusion steps the density of points sampled will be greater. Therefore the dimension will appear to decrease on smaller distance scales. 


In fact, one can actually map the path a fictitious diffusing particle traces out in a given ensemble of triangulations defined by the CDT approach to quantum gravity. In CDT one approximates a continuous spacetime manifold by connecting adjacent 4-dimensional simplices via their mutual tetrahedra, forming a discretised simplicial geometry. The resulting ensemble of triangulations can be used to study different geometric properties of the simplicial manifold, one of which is the spectral dimension, which can be calculated by studying how a test particle diffuses within the geometry. The test particle starts in a randomly chosen simplex and diffuses throughout the ensemble by jumping to adjacent simplices, making $\sigma$ diffusion steps in total. Individual diffusion processes on a triangulation are typically averaged over in order to determine the spectral dimension of the simplicial manifold. However, such diffusion processes might also be used to give information about how the length of such trajectories vary with distance scale, and hence allow one to define an effective velocity of the diffusing test particle. For combinatorially unique triangulations, each simplex the test particle visits during its random walk has a unique label, allowing one to track the diffusion process along each step of its trajectory. 

\begin{figure}[H]
\centering
\includegraphics[width=0.5\linewidth,natwidth=610,natheight=642]{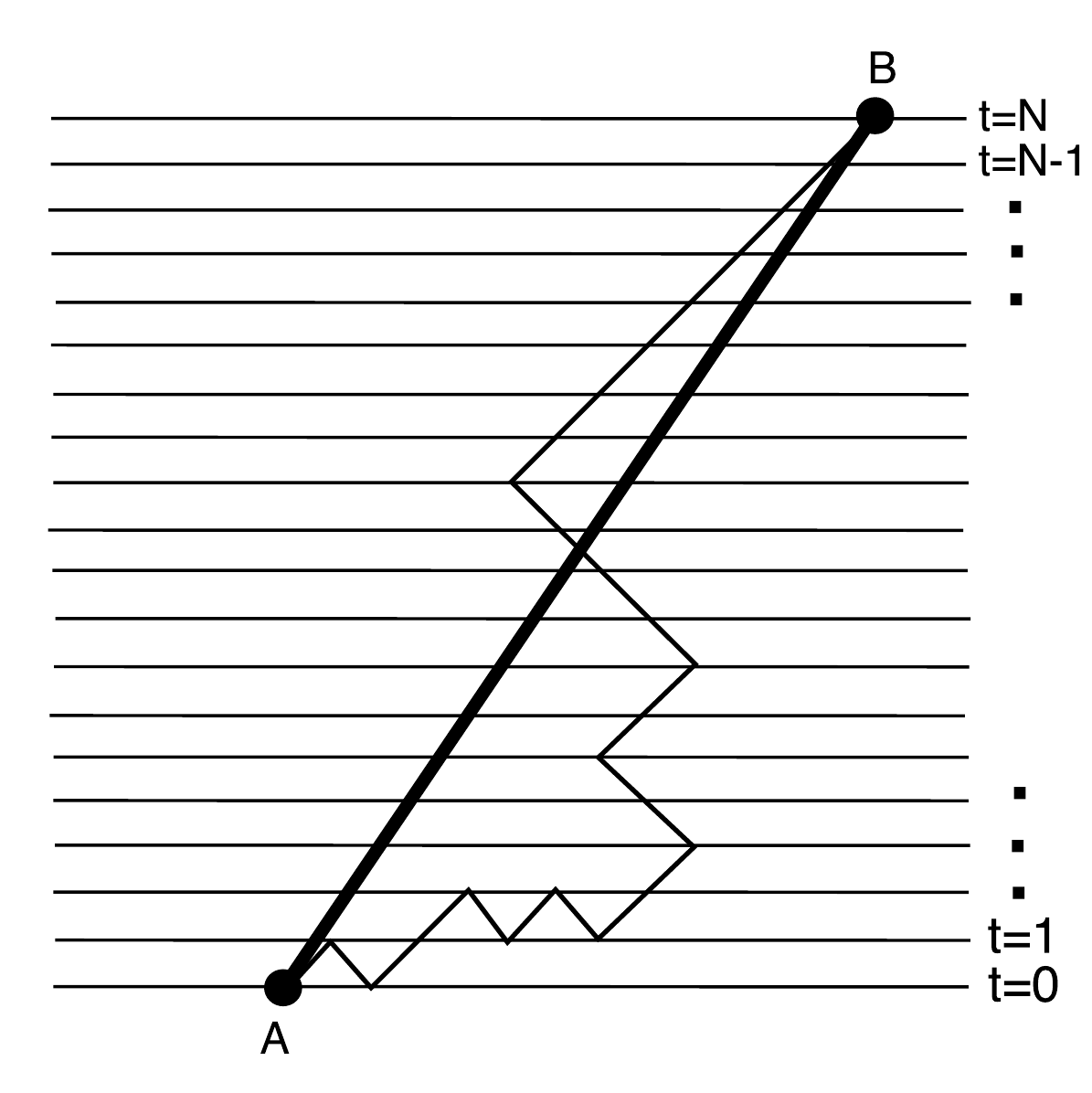}
\caption{\small A schematic representation of a trajectory that a fictitious particle diffusing between points A and B in $(1+1)$-dimensional CDT might take.}
\label{SOLschem}
\end{figure}

A defining feature of CDT is that it distinguishes between space-like and time-like links on the lattice so that an explicit foliation of the lattice into space-like hypersurfaces of fixed topology can be introduced. The space-like hypersurfaces are separated by time intervals $t_{N}$, as shown schematically in Fig. \ref{SOLschem}, thereby explicitly introducing a time coordinate. In CDT the time-like and space-like simplicial edge lengths, $a_{t}$ and $a_{s}$, respectively, do not have to be equal. The path length a diffusing particle traces out within the ensemble of triangulations is just equal to the number of diffusion steps taken, $\sigma$, multiplied by the average distance between adjacent simplices, which we encode by the constant of proportionality $C$, which is a function of $a_{t}$ and $a_{s}$. A CDT triangulation defines a time coordinate that exists within the triangulation at times $t=0$, $t=1$, ..., $t=N$, as shown schematically in Fig. \ref{SOLschem}, defining a causal slice of spacetime of duration $t=N$. The elapsed time observed by a particle diffusing between points A and B is then given by the number of times the test particle crosses a space-like hypersurface, thus incrementing its time coordinate $t$. The number of times the particle's trajectory intersects a space-like hypersurface we denote by $t_{d}$. Hence, we can now define an effective velocity $v_{d}$ for the massless diffusing test particle within the triangulation via

\begin{equation}
v_{d}=\frac{C\sigma}{t_{d}}.
\end{equation}

\begin{figure}[H]
\centering
\includegraphics[width=0.7\linewidth,natwidth=610,natheight=642]{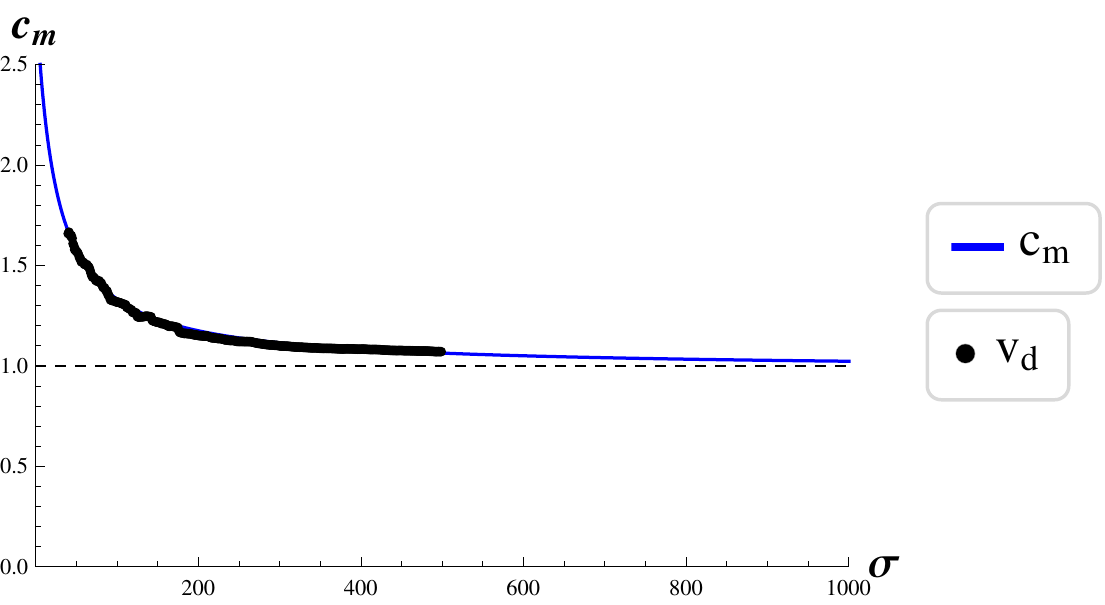}
\caption{\small A comparison between the modified speed of light $c_{m}$ predicted from dimensional reduction (the solid blue curve) and the effective velocity $v_{d}$ determined by averaging over 1000 different diffusion processes in a typical CDT ensemble of triangulations (filled black dots). The dashed black line indicates the speed of light $c=1$.}
\label{SOLdiff}
\end{figure}

Figure \ref{SOLdiff} shows $v_{d}$ averaged over 1000 different diffusion processes for the canonical point in the de Sitter phase of CDT, and for a constant of proportionality $C=0.18$, chosen such that $v_{d}=1$ in the large distance limit, as one would expect of a massless diffusing particle. The important feature of Fig. \ref{SOLdiff} is that the measured velocity of the diffusing particle $v_{d}$ in a typical ensemble of triangulations in CDT closely matches the scale dependent speed of light of Eq. (\ref{cm}).\interfootnotelinepenalty=10000 \footnote{\scriptsize Since there is no speed of light in Euclidean signature (in which the spectral dimension is almost universally studied) it is important to define an effective velocity $v_{d}$, which can instead be thought of as the speed of information transfer.} Figure \ref{SOLdiff} therefore provides numerical evidence for superluminality on small distance scales within the CDT approach to quantum gravity. Based on this numerical evidence it is tempting to conclude that Lorentz invariance in CDT is broken on small distance scales, effectively yielding a scenario along the lines of Ho{\v r}ava-Lifshitz gravity in which space and time scale differently. Indeed, many similarities between CDT and Ho{\v r}ava-Lifshitz gravity have been reported in the literature \cite{Ambjorn:2010hu,Anderson:2011bj,Ambjorn:2013joa}. If this correspondence between the two theories holds, it suggests the effective action of CDT may contain higher spatial derivatives, yielding a description of dimensional reduction in CDT similar to Eq. (\ref{hausSpeed}), which was originally derived within the context of Ho{\v r}ava-Lifshitz gravity.    


In Ref. \cite{AbbottWise} Abbott and Wise examine the quantum mechanical trajectory of a particle by measuring its position with an increasingly refined resolution $\Delta x$, finding that the length of the quantum mechanical path increases as one increases the resolution with which we probe spacetime. Note this property is also a general feature of fractal curves \cite{AbbottWise}. We argue the same thing is happening for the diffusing particle that defines the spectral dimension. As we resolve spacetime on ever decreasing scales the path length of the diffusing particle increases in length according to $\Gamma$, hence we obtain a modified speed of light $c_{m}$, as shown in Fig. (\ref{Gamma3}) and supported by the numerical evidence presented in Fig. \ref{SOLdiff}, given by

\begin{equation}
c_{m}=\frac{\Gamma \Delta l}{\Delta t}.
\end{equation}

The fictitious particles diffusing within the CDT ensemble of triangulations are massless and so should propagate at the speed of light. Now, if the massless diffusing particle is a photon bouncing between two parallel mirrors, the photon's path length will appear to get longer as an observer probes the trajectory with ever finner measurements. If we set up a light-clock such that each time the photon traverses the distance between the two mirrors we define the tick of a clock, and if we are to assume the constancy of the speed of light then the observer must see the light-clock ticking slower than it would if probed on larger distance scales. We argue the same thing is true of diffusion processes that define the spectral dimension. Since our postulate assumes the speed of light is scale invariant, and since the path length increases by a factor $\Gamma$ as a function of $\sigma$, time must also dilate on smaller distances by the same $\Gamma$ factor, such that we obtain a scale invariant speed of light $c$, hence

\begin{equation}
c_{m} \rightarrow c =\frac{\Gamma \Delta l}{\Gamma \Delta t}.
\end{equation} 

If we define $\Delta t$ as the time it takes a photon to traverse the distance between the mirrors along the shortest possible path, and $\Delta t'$ as the time it takes the photon to traverse the distance between the two mirrors when an observer probes the trajectory with a resolution of $\Delta x$, we obtain a relation reminiscent of time dilation in special relativity, namely

\begin{equation}
\Delta t'=\Gamma \Delta t.
\end{equation}

\noindent We conclude that dimensional reduction as reported in CDT, and possibly other approaches to quantum gravity exhibiting dimensional reduction, is in fact telling us that in order to retain a scale invariant speed of light time must dilate as a function of distance scale.

\end{section}


\begin{section}{A dual description?}\label{ComplexT}

Following Hausdorff, we can introduce a new definition of length $\langle L \rangle$ that is independent of the measurement resolution $\Delta x$ via a rescaling by the number of spatial Hausdorff dimensions $d_{H}$ \cite{AbbottWise},

\begin{equation}
\langle L \rangle=\langle l \rangle \left(\Delta x\right)^{d_{H}-1}.
\end{equation}

\noindent By defining the diffusion paths in terms of the Hausdorff length $\langle L \rangle$ they are by construction scale invariant, with the scale dependence now encoded in the Hausdorff dimension. The ratio of the invariant Hausdorff length $\langle L \rangle$ and the variable length $\langle l \rangle$ is by definition $1/\Gamma$, where $\Gamma$ is defined by Eq. (\ref{GammaFuncX}), so that

\begin{equation}
\frac{\langle L \rangle}{\langle l \rangle}=\frac{1}{\Gamma}=\left(\Delta x\right)^{d_{H}-1}.
\end{equation}

\noindent Which gives,

\begin{equation}
d_{H}=\frac{\ln{\left(1/\Gamma\right)}}{\ln{\left(\Delta x\right)}}+1.
\label{dhminone}
\end{equation}

\noindent Equation (\ref{dhminone}) describes how the spatial Hausdorff dimension $d_{H}$ of particles diffusing in a typical CDT geometry changes as a function of the distance scale with which they are probed $\Delta x$. 

We now return to Eq. (\ref{sol1}), namely 

\begin{equation}
c_{m}=\frac{E}{p}=\sqrt{1+\left(\lambda p\right)^{2\gamma}}.
\end{equation}

\noindent Recalling the discussion in Section \ref{motivations}, we remind the reader that Eq. (\ref{sol1}) can only be applicable to approaches to quantum gravity that admit the possibility of a preferred global frame of reference, such as Ho{\v r}ava-Lifshitz gravity and possibly CDT. In such illustrative cases we obtain a speed of light $c_{m}$ from Eq. (\ref{sol1}) that is independent of the scale $\lambda p$ by setting $\gamma=0$. For simplicity we can normalise Eq. (\ref{sol1}) so that $c_{m}=1$ when $\gamma=0$, giving

\begin{equation}
c_{m}=\frac{E}{p}=\frac{1}{\sqrt{2}}\sqrt{1+\left(\lambda p\right)^{2\gamma}}.
\end{equation}

\noindent We now plug $\gamma=0$ into Eq. (\ref{hausSpeed}), and upon rearranging we obtain the condition

\begin{equation}
d_{H}=D_{S}-t_{H}.
\label{condition}
\end{equation}

\noindent Combining Eqs. (\ref{dhminone}) and (\ref{condition}), and fixing $D_{S}=2$, tells us that the number of temporal Hausdorff dimensions $t_{H}$ transforms as a function of scale according to

\begin{equation}
t_{H}=1-\frac{\ln\left(1/\Gamma\right)}{\ln\left(\Delta x\right)}.
\label{temp1}
\end{equation}

Figure \ref{hauspath3} shows Eqs. (\ref{dhminone}) and (\ref{temp1}) on a single plot. We see that in the infra-red limit we have $d_{H}=1$ and $t_{H}=1$, whereas in the ultraviolet limit we have $d_{H}\rightarrow 0$ and $t_{H}\rightarrow 2$. The CDT diffusion paths therefore seem to suggest that a spatial dimension is transforming into a temporal dimension as we decrease the distance scale, which can be interpreted as an inverse Wick rotation. The recent work of Ref. \cite{Ambjorn:2015qja} reports a possible scale dependent signature change of the metric in CDT, which if confirmed would provide numerical evidence for this claim. A Wick rotation has also been reported in loop quantum gravity \cite{Mielczarek:2012pf}. 

A dual description of dimensional reduction of the spectral dimension in CDT appears to exist: the initially linear $d_{H}=1$ diffusion paths on macroscopic scales dissolve into a series of zero-dimensional points on microscopic scales, while $t_{H}$ simultaneously increases from $t_{H}=1$ to $t_{H}=2$. Such an explanation is similar to results reported in \cite{Nicolini:2010dj} and is somewhat reminiscent of asymptotic silence. Clearly, if we are to restore $d_{H}=1$ and $t_{H}=1$ on all distance scales then we must reverse this process, transforming a dimension of time back into a dimension of space as a function of distance scale, i.e. perform a scale dependent Wick rotation $t\rightarrow -i\tau$ in order to once again recover a scale invariant speed of light.

\begin{figure}[H]
\centering
\includegraphics[width=0.7\linewidth,natwidth=610,natheight=642]{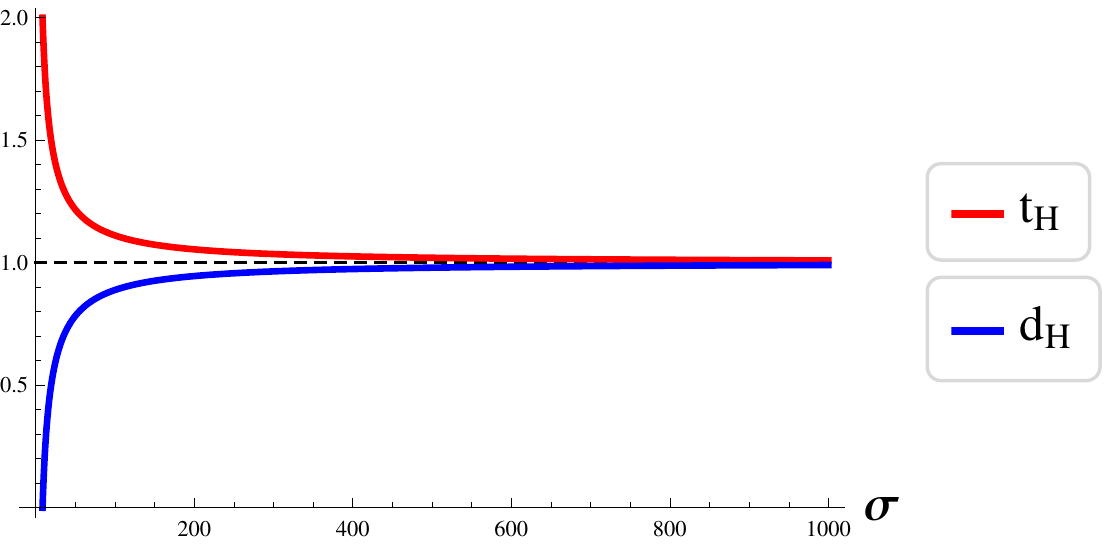}
\caption{\small The spatial $d_{H}$ and temporal $t_{H}$ Hausdorff dimensions of CDT diffusion paths as a function of distance scale $\sigma$ using the $\Gamma$ function of Eq. (\ref{GammaFuncSig}). Note that $d_{H}+t_{H}=2$ over all distance scales.}
\label{hauspath3}
\end{figure}

\end{section}


\begin{section}{Discussion and conclusions}

Combining key elements of general relativity and quantum mechanics predicts a particular property of spacetime known as quantum foam. The energy-time uncertainty relation coupled with mass-energy equivalence implies that as one probes spacetime on Planckian scales one encounters an extremely turbulent geometry, with the possibility of creating infinite energy fluctuations as the distance scale is taken to zero. Such fluctuations of spacetime on small distance scales can lead to large scale observable effects, such as the energy dependence of the speed of light in a vacuum. Likewise, many attempts to combine general relativity and quantum mechanics in a more rigorous mathematical framework predict a similar dispersion of light in a vacuum. However, experiments performed by the Fermi space telescope and others have already measured the difference in arrival times for photons with different energies emitted from the same cosmic source \cite{Nemiroff:2011fk}. The idea being that shorter wavelength photons resolve the turbulent geometry to a greater extent than longer wavelength photons, thus comparatively delaying their time of flight. The Fermi GBM/LAT collaboration using the Fermi Gamma-ray space telescope have put such severe constraints on the existence of quantum foam, even at energies exceeding the Planck scale for dispersion relations that linearly depend on energy, as to effectively rule out such phenomena \cite{FermiGBMLAT1}. 

If the speed of light were linearly dependent on the energy scale, or conversely on the distance scale, then such an effect would most likely already have been observed by the Fermi collaboration. Since this is not the case, the dispersion relation for light in a vacuum is likely either dependent on some higher power of the energy scale, or alternatively entirely scale independent. Constraints on dispersion relations that have a higher power energy dependence are of course currently much less restrictive, and non-systematic Lorentz violations or certain deformations of the relativistic symmetries remain almost entirely unconstrained. However, the fact that the Fermi results are just another example of empirical data that has proven consistent with Lorentz invariance being an exact symmetry of nature is at the very least suggestive. 

In a similar manner to the way that experimental attempts to measure the aether guided the development of special relativity, it seems that experimental observations are just starting to reach a sensitivity to be useful in the development of quantum gravity. In fact it may already be the case that the prediction of a scale dependent speed of light in CDT quantum gravity is at odds with astronomical observations \cite{Mielczarek:2015cja}. Wherever such possible conflicts exist, and wherever it makes sense to do so, we propose that approaches to quantum gravity might be made to conform with experiment via the inclusion of a scale dependent time coordinate, such that the speed of light always remains a scale invariant quantity. 

  

This work provides numerical evidence for a scale dependent speed of light within CDT quantum gravity, a property that is also present in a number of other approaches such as Ho{\v r}ava-Lifshitz gravity \cite{Horava:2009uw}, loop quantum gravity \cite{Smolin:2004sx} and noncommutative geometry \cite{Cai:2001az}. We propose that if we are to maintain a scale invariant speed of light in approaches to quantum gravity that predict superluminality then we must admit a modification to the nature of time; time must dilate as a function of relative scale, just as it does as a function of relative velocity. The inclusion of such a modification to the nature of time may also have important implications for the perturbative renormalizability of gravity, since in this scenario the zero distance limit may no longer correspond to the infinite energy limit. By using the Hausdorff definition of invariant length we determine the spatial and temporal Hausdorff dimensions of diffusing particles within CDT geometries, finding an effective scale dependent signature change of the metric on small distance scales, thus providing a possible dual description of dimensional reduction in CDT.

\end{section}


\section*{Acknowledgements}

Firstly, I would like to express my gratitude to Jerzy Jurkiewicz for encouraging discussions, useful comments and for sharing the manuscript with his colleagues. I'm also indebted to Jan Ambjorn, Gordon Donald and Jack Laiho for their comments on the manuscript, and to Antonio Morais for early discussions. I wish to acknowledge the support of the grant DEC-2012/06/A/ST2/00389 from the National Science Centre Poland. 

\begin{section}{References}

\bibliographystyle{unsrt}
\bibliography{Master}

\end{section}

\end{document}